# What the difference between Thymine and Uracil?

Semenov D.A. (dasem@mail.ru)

**Abstract.** The wobble hypothesis does not discriminate between uracil and thymine. Methylation could favor further stabilization of uracil in the keto form. Thymine is present in keto form only and can pair up but with adenine. Uracil can easily construct the enol form; that is why it forms the U-G pair.

In previous E-print I proposed an alternative to the wobble hypothesis [1]. According to my reviewers it is "keto-enol hypothesis" to form G-U pair.

Confirmation of my hypothesis can be found in studies systematically addressing uracil derivatives in the wobble position.

Incorporation of thymine into DNA may be accounted for as follows. Methylation could favor further stabilization of uracil in the keto form. Oxygen is an electron acceptor while the methyl group an electron donor, so the presence of the methyl group reduces the probability of the proton being near oxygen. The wobble hypothesis does not discriminate between uracil and thymine, so it can be tested in the experiment with thymine substituted for uracil in the codon. It is not difficult to paste the thymine in to mRNA.

The influence of the derivative in the 5' position of uracil on the formation of the enol form was addressed in several studies: Sowers and colleagues [2] made the conclusion that the mutagenic activity of 5'-bromouracil is related to the relative ease of the formation of the enol form.

Thymine and uracil are considered to be nucleotides with similar properties, while 5'-bromouracil is assumed to be different from both of them. The fact that the difference between thymine and uracil in their ability to make enol form is consistently ignored prevents one from realizing that DNA contains thymine and RNA – uracil. My interpretation of this is as follows: thymine is present in keto form only and can pair up but with adenine. Uracil, on the other hand, can easily construct the enol form; that is why it forms the U-G pair and for the same reason it would lead to mutagenesis, like 5'-bromouracil, if it were contained in DNA.

Zeegers-Huyskens [3] theoretically demonstrated the influence of derivatives in the 5' position, confirming that the presence of the electron-accepting derivative decreases the basicity of oxygen in the 4' position. Of special interest is the result for thymine: the donor derivative (methyl group) qualitatively changes the result – the basicity of oxygen increases dramatically. That is, thymine has greater difficulty to construct enol form. This result places uracil in one group with its 5' substituted derivatives, if the derivative is an acceptor of σ-electron density, while thymine drops out of this group.

Realization of this fact makes it possible to interpret three results immediately related to the wobble hypothesis:

Takai and his co-authors [4] demonstrated that 5'-oximetyluridine in the wobble position facilitates the formation of the wobble pair.

Näsvall and colleagues [5] state that even 5'-oxiacetyluridine facilitates the formation of the wobble pair.

According to Kurata and colleagues [6], the same effect was produced by 5-carboxymethylaminomethyluridine and 5-taurinomethyluridine, which contradicted the expectations of the experimenters. In each case, a new substantiation has to be invented for the results.

One can avoid this constant violation of the principle of reasoning named Occam's razor by recognizing that the U-G pair has been made by the enol form of uracil. Then, all the above-mentioned experimental data can be naturally interpreted as the effect of the acceptor derivative in the 5' position of uracil.

Moreover, Takai and Yokoyama [7] demonstrated the absence of the proton in the 3' position of uracil under similar conditions, which is a direct proof of my hypothesis.

**Acknowledgement.** The author would like to thank Krasova E. for her assistance in preparing this manuscript.

**References:**

1. Semenov D.A. Wobbling of What? arXiv:0808.1780
2. Sowers, L.C; Goodman, M.F; Eritja, R; Kaplan, B; Fazakerley, G.V. Ionized and wobble base-pairing for bromouracil-guanine in equilibrium under physiological conditions. A nuclear magnetic resonance study on an oligonucleotide containing a bromouracil-guanine base-pair as a function of pH. *J Mol Biol.* 1989;**205**:437–447
3. Zeegers-Huyskens Th. The basity of the two carbonil bonds in uracil derivates. J. Mol. Struct. 198. 135-142. (1989)
4. Takai K ., Okumura S., Hosono K., Yakoyama S., Takaku H. A single uridine modification at the wobble position of an artificial tRNA enhances wobbling in an Escherichia coli cell-free translation system . FEBS Letters , 447 , 1 , 1 - 4 (1999)
5. Näsvall S. J., Chen P., Björk G.R. The wobble hypothesis revisited: Uridine-5-oxyacetic acid is critical for reading of G-ending codons. RNA, 2007; 13(12): 2151 - 2164
6. Kurata S, Weixlbaumer A, Ohtsuki T, Shimazaki T, Wada T, Kirino Y, Takai K, Watanabe K, Ramakrishnan V. and Suzuki T Modified uridines with C5-methylene substituents at the first position of the tRNA anticodon stabilize U·G wobble pairing during decoding J. Biol. Chem. 2008 283(27): 18801-11
7. Takai, K., and Yokoyama, S. Roles of 5-substituents of tRNA wobble uridines in the recognition of purine-ending codons (2003) *Nucleic Acids Res* **31**(22), 6383-6391